\documentclass[12pt]{article}
\usepackage{multirow}
\usepackage{amsmath,amsfonts,amssymb,bm,slashed,bbm,mathrsfs}
\usepackage{graphicx,color}
\usepackage{graphicx}
\graphicspath{{./Pic/}}
\renewcommand{\Ref}[1]{(\ref{#1})}
\newcommand{\eq}[2]{\begin{align}\label{#1}#2\end{align}}

\newcommand{\nn}{\nonumber}

\newcommand{\pa}{\partial}
\newcommand{\ep}{\epsilon}

\newcommand{\sig}{\sigma}

\newcommand{\la}{\lambda}

\renewcommand{\Ref}[1]{(\ref{#1})}
\newcommand{\beao}{\begin{eqnarray*}}
\newcommand{\eeao}{\end{eqnarray*}}
\newcommand{\be}{\begin{equation}}
\newcommand{\ee}{\end{equation}}
\newcommand{\bea}{\begin{eqnarray}}
\newcommand{\eea}{\end{eqnarray}}
\newcommand{\beq}{\begin{eqnarray}}
\newcommand{\eeq}{\end{eqnarray}}%vesion with corrections 070221
\begin{document}
%\begin{center}
\title{Spontaneous magnetization of QGP at high temperature}
\author{
  V. Skalozub\thanks{e-mail: Skalozub@dnu.edu.ua} \\
{\small Oles Honchar Dnipro National University, 49010 Dnipro, Ukraine}}
\date{ }
%I. Gamolsky \\
%{\small Oles Honchar Dnipro National University, 49010 Dnipro, Ukraine}
%\end{center}
\date{\small}
%\title{ Quark propagation at  Polyakov's loop  background}
%\author{V. Skalozub,  A. Turinov}
%{\small Oles Honchar Dnipro National University, 49010 Dnipro, Ukraine}}
%\affil{Dnipro National University, Dnipro, Ukraine}
%\setcounter{Maxaffil}{Dnipro National University, Dnipro, Ukraine}
%\renewcommand\Affilfont{\itshape\small}
\maketitle
%%%%%%%%%%%%%%%%%%%%%%%%%%%%%%%%%%%%%%%%%%%%%%%%%%%%%%%%%%%%%%%%%
%\begin{center}
%{\bfseries
\begin{abstract}
In quark-gluon plasma (QGP), at  higher  deconfinement temperatures $T \ge  T_d$ the spontaneous generation of color magnetic fields, $b^3(T), b^8(T)  \not = 0$ (3, 8 are color indexes), and usual magnetic field $b(T)  \not = 0$ happens. Simultaneously,   the  Polyakov loop  and/or  algebraically  related to it $A_0(T)$ condensate, which is solution to Yang-Mills imaginary time equations,    are also created.
Usually, in analytic quantum field theory  these effects are investigated independently of each other within the effective potentials having different mathematical structures.
The common generation of these condensates  was detected   in  lattice Monte Carlo simulations.

Recently, with  the new type two-loop effective potential, which generalizes  the known integral representation for the  Bernoulli  polynomials and  takes into consideration the magnetic background, this effect   has been  derived analytically.
  The corresponding      effective potential $W(T, b^3, A_0 )$  was  investigated  either in SU(2) gluodynamics or full QCD. The gauge fixing independence of it was   proved within the Nielsen identity approach. The values of magnetic field strengths at different temperatures were calculated  and the mechanism of stabilizing   fields  due to $A_0(T)$ condensate has been  discovered. In the present review, we describe this  important phenomenon in more details,  as well as a number of specific  effects - induced color charges, effective photon-photon-gluon vertexes - happening due to vacuum polarization at this background.   They could serve  as the    signals of the QGP creation in the heavy ion collision experiments.
	
Key words: spontaneous magnetization, high temperature, asymptotic freedom, effective potential, $A_0$ condensate, effective charge, effective vertexes.
\end{abstract}
%\end{center}
\section{Introduction}
Deconfinement phase transition (DPT),  as well as the properties of the quark-gluon plasma (QGP), are widely investigated  for many years. Most results have been obtained in the lattice simulations because of the large coupling value $g \ge 1$ at the  phase transition temperature $T_c$. But at high temperatures due to asymptotic freedom the analytic methods are also reliable. They give a possibility for investigating  various phenomena in the plasma. Among them is the creation of gauge field condensates described by the classical solutions to field equations without sources. Only such type  fields could appear spontaneously inside the QGP. The well known ones are the so-called $A_0$ condensate, which is algebraically related to the Polyakov loop (PL) and  the chromomagnetic fields $b^3 = g H^3, b^8 = g H^8$ (3, 8 are color indexes of SU(3) group)  which are the Savvidy vacuum states at high temperature. These condensates result in numerous proper new effects  which could be the signals of the QGP. The  condensation of   $A_0$ alone is investigated by  different methods. For  recent works see, for instance, \cite{gaof21-103-094013} and references  therein.

All the mentioned condensates are the consequences of asymptotic freedom and follow from the important property  that asymptotic freedom at high temperature inevitably  results in an infrared instability at low one. The field condensation prevents such type instability that results in the formation of the physical  vacuum state.  In quantum field theory (QFT), the magnetic and the $A_0$ condensates are generated at different orders  in coupling constant (or the  number of loops) for the effective potential (EP) $W(T, b, b^3, b^8, A_0)$. So that they have different temperature dependencies and  play different  roles in the QGP dynamics.   For example, $A_0$ is generated at $g^4$ order in coupling constant and determined by the ratio of two- and one-loop  contributions to $W(A_0)$. So it has  the $g^2$ order. The fields $b(T), b^3(T), b^8(T)$ are generated in tree - plus one-loop - plus daisy approximation and also have the order $g^2$ in coupling constant. But they have  other temperature dependence due the contribution to the EP of  the tree-level term coming from the classical  equation   solutions. On the other hand, the contribution of $A_0$ at tree level equals  zero because it is a constant electrostatic potential. This difference   is    important at high temperature. All mentioned features  require special comprehensive  considerations.

The fields investigated below are an important topic towards a theory of confinement. The $A_0$-background is relevant because at finite temperature such field cannot be gauged away and is intensively investigated beginning with \cite{weis82-25-2667}. In the early 90-ies, two-loop contributions were calculated in QCD and with these,  the EP has  non-trivial minimums and related condensate fields (see, for instance, \cite{skal94-57-324}, \cite{skal21-18-738}). They form a hexagonal structure in the plane of the  color components $A_0^3$ and $A_0^8$ of the background field.

The other kind  background is the chromomagnetic one. More details about this field and the ways of its stabilization at finite temperature can be found, in particular,  in  \cite{skal00-576-430}, \cite{demc15-46-5},\cite{skal21-mon} . The magnetization  is also resulted from the minimum of the EP, which is stable in the consistent approximation of one loop plus daisy diagram contributions. In the review  \cite{demc15-46-5} and  book \cite{skal21-mon} the results of different approaches (analytic (in noted approximation) and numeric) are presented, in particular, on lattice calculations with $A_0$ and chromomagnetic field.

 A common generation of both fields was studied analytically in \cite{bord22-82-390}. Here, new  representation  generalizing the known integral representation for the Bernoulli polynomials, was worked out, which  admits introducing either $A_0$ or any $b$ fields up to two-loop order. Below we write $b$ for each magnetic field, for brevity. Within this representation, in particular, the known results for separate generation of the fields have been reproduced. However, the spontaneous generation of chromomagnetic field up to two-loop order was not investigated in detail. So, the mechanism of the vacuum stabilization remained not clarified finally.

 This  problem was analytically investigated in \cite{skal22-30(1)-3}. It   is of grate importance because in the lattice calculations accounting for both backgrounds  \cite{demc13-21-13} it has been observed that in the presence of the constant color magnetic field the PL acquires  a non-trivial spatial structure    along the  direction of the field. More interesting,  on the lattice also,  a common spontaneous generation of both fields was detected  \cite{demc08-41-165051}, \cite{skal21-mon}. So, to clarify a mechanism of magnetic field stabilization in QFT taking into consideration both condensates, one has to turn to two-loop calculations. This is because    the stabilization of magnetic field within the one-loop plus daisy diagrams does not work. Such  approximation is insufficient in case of two fields. This is because the generation of the latter one is realized at two-loop level for the EP. In what follows, we discuss in details both these approximations and compare the obtained results. In fact, in the current literature the cases of the $A_0$ and $b$ condensations are investigated separately. So, mutual relations of them remained not clarified and estimated qualitatively, only.

 In what follows, first, we calculate the EP as the function of  $A_0$, and $b = g H^3$, in $SU(2)$ gluodynamics. The extension to full QCD is trivial because it includes three such groups.  The integral expressions for the EP of the $A_0$  are generalized  to include the magnetic background.
Also, we consider the limiting cases $A_0=0$ and $b \not =0$ and find, for instance, the magnetic condensate in two-loop order, which was also considered in  \cite{skal00-576-430}, \cite{bord22-82-390}  but using other approaches and in not wide temperature interval.

Note again that the spontaneous generation  of a background field is  meant in the sense  that for the corresponding field the EP has a minimum below zero, which is energetically favorable.
In QGP, the $A_0(T)$ ( or/and PL)  results  in the color $Z(3)$ symmetry breaking and the Furry theorem violation. The magnetic fields considerably change the spectra of quarks and gluons as well.

 So, new phenomena have to be realized. In particular, the induced color charges $Q^3_{ind}, Q^8_{ind}$ and the physically unexpected new type vertexes joining photon and gluon states could be generated. Obviously that at low temperature this is impossible, the white states and the colored ones do not interact (unite)  in one vertex.  The PL as well as  $A_0(T)$ are the order parameters for the DPT.    At low temperature they equal zero. At high temperature they become nonzero. The same concerns the spontaneously created magnetic fields.  Recently, on the principles of the Nielsen's identity method and new type integral-sum representation for the EP   we derived the gauge invariant expression for the $ A_0$ condensate in the magnetic fields in two-loop approximation, \cite{bord22-82-390}, \cite{skal21-18-738}. This (in particular)  opens a possibility for calculating the induced color charges and other effective vertexes  for this general background of  QGP.
% In particular, the induced color charges $Q^3_{ind}, Q^8_{ind}$ and the physically unexpected new type vertexes %relating photon and gluon states could be generated.

  %Obviously that at low temperature this is impossible, the white states and the colored ones do not interact (relate)  in one vertex.  The PL as well as  $A_0(T)$ are the order parameters for the deconfinement phase transition.    At low temperature, PL and $ A_0$ equal zero. At high temperature they become nonzero. The same concerns the magnetic fields.  Recently, on the principles of the Nielsen's identity method and new type integral-sum representation (ISR)   we derived the gauge invariant expression for the $ A_0$ condensate in the magnetic fields in two-loop approximation, \cite{bord22-82-390}, \cite{skal21-18-738}. This (in particular)  opens a possibility for calculating the induced color charges and other effective vertexes  for this general background of the QGP.

  To realize that, we have  to calculate the contribution of  diagrams depicted in Fig. 1 and Fig. 2.  There in, the solid line presents the quark propagator in the $A_0$ and magnetic fields and the wavy line presents the zero component of gluon fields $ G_0^3$ or $G_0^8$. At finite temperature $T$, in the Matsubara formalism, one has to calculate the temperature sum  over discrete energy values  $p_4 = 2 \pi T (l + 1/2), l = 0, \pm 1,  ...$, integrate over momentum component $p_3$ oriented along the space field direction, calculate the sum over spin variable  $\sigma = \pm 1$  and  sum up over n = 0, 1, 2,..., in correspondence to the fermion spectrum in magnetic field b: $(p_4 + g A_0)^2 = m^2 + p_3^2 + (2 n + 1)  b - \sigma  b $. Here we write  $ b$ as a general expression corresponding to each of the fields.  This is $e H$ for usual magnetic field, $g H^3, g H^8$ - for color fields.

  In actual calculations of investigated effects, we apply  the low level approximation, $n = 0, \sigma = + 1$ giving a leading contribution for strong external fields.
We  obtain that the induced color charge $Q^3_{ind}$ is nonzero. The presence of the magnetic fields changes the values of it compared to the zero field case. As a result, we derive that  QGP has to be magnetized and  color charged.

The way of presenting the results is as follows. First, in sect. 2 we introduce and discuss the general two-loop EP of both field calculated in the background $R^{ext.}_\xi$ gauge. This EP  will be investigated for various cases of interest. So, one is able to proceed further taking into consideration only this one. However, for more detailed  considerations,  in sect. 3  we present the results on the Nielsen identity method for proving the gauge fixing parameter ($\xi$) independence of the EP. In sect. 4 we find the relation for the initial $\xi$-dependent EP and the EP of order parameter PL \cite{skal21-18-738}. Then in sect. 5 we consider the case of $b$  field generation. The case of $A_0$ only is additionally investigated in sect. 6.  This section closes  describing the creation of the stable  QGP background  at high temperature. In next sections we consider new type vertexes and  related with them effects which have to happen  and  manifest the creation of QGP. Namely, in sect. 7 we calculate the induced color charge $Q_{ind.}^3$ followed from diagram in Fig. 1. In sect. 8 we calculate and investigate the effective vertex in Fig. 2 which joins  one gluon and two photon states. A number of effects related with this vertex is discussed.

\begin{figure}[h]
\begin{center}
\includegraphics{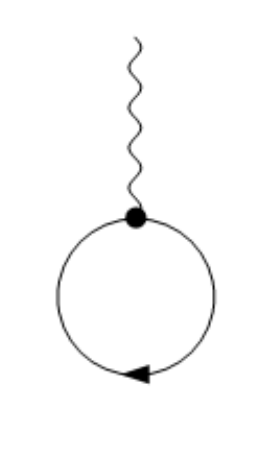}
\caption{Tadpole diagram}
\label{tadpole}
\end{center}
\end{figure}
%\begin{figure}[h]
%\begin{center}
%\includegraphics[width=0.25\textwidth]{ql2.png}
%\caption{Tadpole diagram}
%\label{tadpole}
%\end{center}
%\end{figure}

\begin{figure}[h!]
\centering
\includegraphics{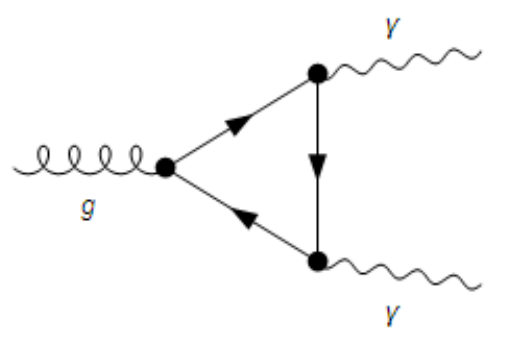}
\caption{Effective $\gamma-\gamma- G^3 (g)$ diagram  }
\end{figure}

\section{Effective potential of fields}
In  the case of SU(2), the effective potential in the background $R_\xi$ gauge reads \cite{bord22-82-390}:
\eq{W2}{ W^{SU(2)}_{gl} &=B_4(0,0)+2B_4\left(a,b\right)
\\\nn&~~~	+2{g^2}\left[
	B_2\left(a,b\right)^2
	+2 B_2\left(0,b\right) B_2\left(a,b\right)	\right]
	-4{g^2}(1-\xi) B_3\left(a,b\right)B_1\left(a,b\right)
}
with the notation
\eq{ab}{a=\frac{x}{2}=\frac{g A_0}{2\pi T},~~~b=gH_3^3.
}
The chomomagnetic field is directed along third directions in coordinate and color spaces.
Since we work at finite temperature, $ W_{gl}$ is equivalent to the free energy.

The functions $B_n(a,b)$ are defined by
\eq{3}{ B_4(a,b) &= T\sum_\ell\int\frac{dk_3}{2\pi}\frac{b}{4\pi}\sum_{n,\sig}
	\ln\left(\left(2\pi T(\ell+a)\right)^2+k_3^2+b(2n+1+\sig-i0)\right),
\\\nn
B_3(a,b) &=
T\sum_\ell\int\frac{dk_3}{2\pi}\frac{b}{4\pi}\sum_{n,\sig}
\frac{\ell+a}{\left(2\pi T(\ell+a)\right)^2+k_3^2+b(2n+1+\sig-i0)}
\\\nn
	 B_2(a,b) &= T\sum_\ell\int\frac{dk_3}{2\pi}\frac{b}{4\pi}\sum_{n,\sig}
	\frac{1}{\left(2\pi T(\ell+a)\right)^2+k_3^2+b(2n+1+\sig-i0)},
\\\nn		
	B_1(a,b) &=
	T\sum_\ell\int\frac{dk_3}{2\pi}\frac{b}{4\pi}\sum_{n,\sig}
	\frac{\ell+a}{\left(\left(2\pi T(\ell+a)\right)^2+k_3^2+b(2n+1+\sig-i0)\right)^2}.
}
In eq.\Ref{W2}, $\xi$ is gauge fixing parameter,  the summations run  $n=0,1,\dots$, $\sig=\pm2$ and $\ell$ runs over all integers. The $'-i0'$-prescription  defines the sign of the imaginary part for the tachyon  mode.
These formulas and eq.\Ref{W2} are the generalization of the corresponding two-loop expressions in \cite{enqv90-47-291}, eqs.(3.8) and (A.2)-(A.5), \cite{bely91-254-153}, eq.(14),  \cite{skal92-7-2895}, eq.(4),  and also \cite{skal21-18-738}, eq.(4),
to the inclusion of the magnetic field. Note   the sign "-" in eq.\Ref{1.1}. Below we  use also the relations
\eq{3a}{B_3(a,b) &= \frac{1}{4\pi T}\pa_a B_4(a,b),
	~~~&	B_1(a,b) &= \frac{-1}{4\pi T}\pa_a B_2(a,b).
}

For $b = 0$ we have to replace  $\frac{b}{4\pi}\sum_{n,\sig}\to\int\frac{d^2k}{(2\pi)^2}$ and get
\eq{1.1}{ 	&&B_4(a,0)=\frac{2\pi^2 T^4}{3} B_4(a),
	 	~~	B_3(a,0)=\frac{2\pi T^3}{3}	B_3(a), \\ \nonumber
  		 &&B_2(a,0)=\frac{T^2}{2}	B_2(a),
	~~~~~~~	B_1(a,0)=-\frac{ T}{4\pi B_1(a)},
}
where $B_n(a)$ are the Bernoulli polynomials, periodically continued.
The special values for, in addition, $a=0$ are
\eq{4}{ B_4(0,0)  = -\frac{\pi^2T^4}{45},
	~~~   B_3(0,0)= 0,
	~~~   B_2(0,0)= \frac{T^2}{12},
	~~~   B_1(0,0)= \frac{T}{8\pi}.
}
We note that these formulas hold for $T>0$.
The motivation for the above choice of the notations  is that  the functions $B_n(a,b)$, \Ref{3}, are  the corresponding mode sums without additional factors. More details about this representation as well as the renormalization and the case of $T = 0$ are given in \cite{bord22-82-390}.
\section{Effective potential and Nielsen's identity}
Now,  let us consider SU(2) gluodynamics in  Euclidean space time,   for the case of zero magnetic field and  nonzero background field $\bar{A}^a_\mu = A_0 \delta_{\mu 0} \delta^{a 3} = const$, described by the Lagrangian
\be \label{Lagr} L = \frac{1}{4} (G^a_{\mu\nu})^2 + \frac{1}{2\xi}[(\bar{D}_\mu A_\mu)^a]^2 - \bar{C} \bar{D}_\mu D_\mu C. \ee
The gauge field potential $A_\mu^a = Q_\mu^a + \bar{A}^a_\mu $ is decomposed in quantum and classical parts. The covariant derivative in  eq.\Ref{Lagr} is $(\bar{D}_\mu A_\mu)^{ab}= \partial_\mu\delta^{ab} - g \epsilon^{abc}\bar{A}^c_\mu,  G^a_{\mu\nu} = (\bar{D}_\mu Q_\nu)^a - ( \bar{D}_\nu Q_\mu)^a - g   \epsilon^{abc} Q^b_\mu  Q^c_\nu, g $   is gauge coupling constant, internal index a = 1,2,3. The Lagrangian of ghost fields $\bar{C}, C$  is determined by the background covariant derivative $  \bar{D}_\mu(\bar{A}) $ and the total one  $ D_\mu(\bar{A} + Q)$. As in  \cite{bely90-45-355}, \cite{bely91-254-153} we introduce the ”charged basis” of fields:
\bea \label{fields}&& A^0_\mu =  A^3_\mu,~~ A^{\pm}_\mu = \frac{1}{\sqrt{2}} (A^1_\mu \pm i A^2_\mu), \\ \nn
&&  C^0 =  C^3,~~ C^{\pm} = \frac{1}{\sqrt{2}} (C^1  \pm i C^2 ). \eea
In this basis a scalar product is $x^a y^a = x^+ y^- + x^- y^+  + x^0 y^0$ , and the structure antisymmetric factors are: $\epsilon^{abc}$ = 1 for a = ”+”, b = ”-”, c = ”0”. Feynman's rules are the usual ones for the theory at finite temperature with modification: in the background field  the sum over frequencies should be replaced by $\sum_{k_0}, k_0 =( \frac{2 \pi l}{\beta} \pm g \bar{A}_0)$ in all loops of the fields $Q_\mu^{\pm}, C^{\pm}$. Here,  $l = 0, \pm 1, \pm 2,...$.  This frequency shift must be done not only in propagators but also in three particle vertexes.

Carrying out standard calculations we obtain the two-loop EP \cite{skal21-18-738}
\bea \label{EP} W(x) &=& W^{(1)}(x)+ W^{(2)}(x),\\ \nn
\beta^4 W^{(1)}(x)&=&  \frac{2}{3} \pi^2 [ B_4(0) + 2 B_4(\frac{x}{2}) ], \\ \nn
\beta^4 W^{(2)}(x)&=&\frac{1}{2} g^2 [  B_2^2(\frac{x}{2})+ 2 B_2(0)) B_2(\frac{x}{2})] +
\frac{2}{3} g^2(1 - \xi) B_3(\frac{x}{2})B_1(\frac{x}{2}),\eea
where $B_i(x)$ are Bernoulli's polynomials defined $ modulo$ 1  adduced in Appendix of paper, $x = \frac{g A_0 \beta}{\pi}, \beta = 1/T$. This expression coincides with calculated already in \cite{bely91-254-153}, \cite{skal92-7-2895}. In what follows we consider the interval $0 \leq x \leq 2$.

Let us investigate the minima of it. We apply an expansion in powers of $ g$ and get
\bea \label{W0} \beta^4 W_{min}& =&  \beta^4 W (0)
-\frac{1}{192\pi^2}(3-\xi)^2g^4, \nn \\
&&x =  g^2 \frac{(3 - \xi)}{8 \pi^2} ,\eea
where the first term is the value at zero field.
Actually, an expansion parameter determined from the ratio of two- and one-loop contributions equals to $\frac{g^2}{8 \pi^2}$, and therefore sufficiently large  coupling values g are permissible.
As we see, both the minimum position and the minimum energy value are gauge-fixing dependent. Hence the gauge invariance of the $A_0$ condensation phenomenon is questionable.

This problem was solved within Nielsen's identity method in \cite{skal92-7-2895}, \cite{skal94-50-1150} for $SU(2)$ and $SU(3)$ gluodynamics and in \cite{skal94-9-4747}, \cite{skal94-57-324} for $ QCD$  with quarks.
Since this approach is important for what follows, we describe it in short here.

In \cite{kobe91-355-1} Nielsen's identity for general type EP has been derived:
\be \label{NI} \delta^{'} W (\phi) = W_{, i} \delta  \chi^i (\bar{\phi}), \ee
which describes a variation of $W(\phi)$ due to variation of the gauge fixing term $F^\alpha (\phi).$  In eq.\Ref{NI} $\phi^i$ is gauge field, $\bar{\phi}^i$ denotes a vacuum value of $\phi^i$, comma after $W$ means variation derivative with respect to corresponding variable. Variation $\delta \chi^i  $ describes changing of field $(\bar{\phi})$  due to special gauge transformation which  compensates  variation of a classical action appearing after variation of gauge-fixing function $F^\alpha(\phi) \to F^\alpha(\phi) + \delta  F^\alpha(\phi)$.

In field theory $\delta \chi^i$ is calculated from equation \cite{kobe91-355-1}:
\be \label{varchi} \delta \chi^i = - \Bigl< D^i_\alpha(\phi) \Delta^\alpha_\beta(\phi)  \delta^{'}  F^\beta(\phi) \Bigr>, \ee
where $\bigl< O(\phi) \bigr> $ denotes functional average of $O(\phi)$. In this expression $D^i_\alpha(\phi)$ is generator of gauge group, $ \Delta^\alpha_\beta(\phi) $ is propagator of ghost fields, $ \delta^{'}  F^\beta(\phi)$ is variation of gauge fixing term.

In our case according eq.\Ref{Lagr} $  \delta^{'}  F^\beta(\phi)  = - \frac{1}{2} (\bar{D}_\mu (\bar{ A}) Q_\mu)^\beta \frac{\delta \xi}{\xi}$,  $ D^i_\alpha$ is covariant derivative. In \cite{skal92-7-2895}, eq.(26), the expression was derived (more details on calculations and discussions for $SU(3)$ case see in \cite{skal94-50-1150}, \cite{skal94-57-324}):
\be \label{varci} \delta \chi^0 = \frac{g}{4 \pi \beta} B_1(\frac{x}{2}) \delta \xi. \ee
Nielsen's identity  for two-loop EP reads
\be \label{NISU2} \frac{ d W}{d \xi} = \frac{\partial W^{(2)}}{\partial \xi} +  \frac{\partial W^{(1)}}{\partial x} \frac{\partial x}{\partial \xi} = 0, \ee
where in the order $\sim g^2$ the derivative $\frac{\partial x}{\partial \xi}$ equals to $\frac{\delta \chi^0}{\delta \xi} \times ( \frac{g \beta}{\pi})$ in  eq.\Ref{varci}. The latter factor comes from definition of $x = \frac{g A_0 \beta}{\pi}$.  Since $W^{(2)}$ has the order $g^2$, and  $W^{(1)}$-  $g^0$,
the eq.\Ref{NISU2} states that $W(x, \xi)$ does not change along the characteristic curve
\be \label{character} x = x' + \frac{g^2}{4 \pi^2} B_1(\frac{x'}{2}) (\xi - \zeta) \ee
in the plain of variables $(x, \xi)$, $\zeta $ is an arbitrary integration constant. Thus, there is the set of  orbits where $W(x')$ is gauge-fixing independent. Along them a variation in $\xi$ is compensated by the  special variation of $x'$.
\section{Effective potential of order parameter }
In this section we, following \cite{bely91-254-153}, express the EP eq.\Ref{EP} in terms of  $\left< L \right> $. We call it "effective potential of  order parameter" $W_L(x_{cl})$.  In SU(2) group, in tree approximation, the PL is expressed in terms of $x$ as follows: $\left< L \right> $ = $\cos (\frac{\pi x}{2})$. This formula can be used to  relate a given value of  PL (or $A_0$) and  classical (observable) condensate value with accounting for radiation corrections: $\left< L \right> =\cos (\frac{\pi x_{cl}}{2}) =  \cos (\frac{\pi x}{2}) + \Delta\left< L \right> $. The quantum correction was calculated in one-loop order (eq. (10) in \cite{bely91-254-153}),
\be \label{deltaPL} \Delta\left< L \right> = - \frac{g^2 \beta \sin (\frac{\pi x}{2})}{4 \pi} \int \frac{  d k  }{k_0^+} \bigl [ \frac{1}{( k_0^+)^2 + \vec{k}^2} + \frac{(\xi - 1)( k_0^+)^2 }{(( k_0^+)^2 + \vec{k}^2)^2 }\bigr], \ee
where the notations are introduced:
\be \int d k  = \int \frac{d^3 k}{(2 \pi )^3} \Bigl ( \frac{1}{\beta} \sum_{l = -\infty}^{\infty}  \Bigr ), ~~k_0^+ = k_0 + g A_0, ~~ k_0 = \frac{2 \pi l}{\beta}. \ee
eq.\Ref{deltaPL}  is crucial for what follows.

%In Ref.\cite{bely91-254-153} the  calculation of  $\Delta\left< L \right>$  was not  presented in detail and only the final expression for the EP  %(Eq.(17))  has been given and analysed. We fill in this gap below.

Obviously that the first  term in eq.\Ref{deltaPL} and the term at $(\xi - 1)$ are positively defined functions and should  have the same signs after integrations.
   The second integral in eq.\Ref{deltaPL} is well known, it is expressed in terms of Bernoulli's polynomials \cite{bely90-45-355}, \cite{skal94-57-324},
\be \label{I2} I_2 = - \frac{(\xi - 1)}{4 \pi \beta} B_1(\frac{x}{2}).\ee
%
%Calculation of the first integral, $I_1$,  we reduce to the previous one.
%Introducing the notation $\tilde{k}^2 =( k_0^+)^2 + \vec{k}^2$,  we write its integrand as follows
%
%\bea \label{t1} Intgd. I_1& =&   \frac{  1  }{k_0^+}  \frac{1}{( k_0^+)^2 + \vec{k}^2} =   \frac{  k_0^+ }{[( k_0^+)^2 + \vec{k}^2]^2}  %\frac{1}{1 - \vec{k}^2/\tilde{k}^2 } \nn\\
%&=& \frac{  k_0^+ }{[( k_0^+)^2 + \vec{k}^2]^2} [ 1 + \sum_{l = 1}^{\infty} (\frac{\vec{k}^2}{\tilde{k}^2})^l ] . \eea
%
% Hence, the first term in   $I_1$ coincides (up to the factor$ (\xi - 1))$ with  $I_2$. The other terms are also positive. So, the sings of  %$I_2$ and  $I_1$ must be the same.
% But this is not the case for  Ref. \cite{bely91-254-153} where the opposite  sign for $I_1$ was obtained.

 Now,   we return  to the initial expression in eq.\Ref{deltaPL} and calculate the first term  by using a standard procedure. This is presented in Appendix of \cite{skal21-18-738} and also in \cite{skal21-mon}. The  result is
\be \label{I111}   I_1  = - \frac{1}{2 \pi \beta} B_1(\frac{x}{2}).\ee
Substituting  $I_1$ and $I_2$ in eq.\Ref{deltaPL}, we obtain finally
\be \label{deltaPLa} \Delta\left< L \right> =  \frac{g^2  \sin (\frac{\pi x}{2})}{16 \pi^2}   B_1(\frac{x}{2} ) (\xi + 1). \ee
Just this formula should be used in order to express    the field $ x$  in terms of "classical observable one", $x_{cl}$.

%Note,  in  Ref.\cite{bely91-254-153}    the  final EP of order parameter  (Eq.(17))  corresponds to   the factor $(\xi - 3)$. But according to  %Eq.\Ref{deltaPL}  the opposite  signs of $I_1$ and $I_2$  are impossible. Factor $( - 3 )$ can be obtained only for positive  sign in %Eqs.\Ref{I111}, \Ref{I11}.

%Summing up these expressions we obtain finally
%
%\be \label{deltaPLa} \Delta\left< L \right> =  \frac{g^2  \sin (\frac{\pi x}{2})}{16 \pi^2}   B_1(\frac{x}{2} ) \xi . \ee
%
%Just this formula should be used in order to express    "nonphysical field"$ x$  in terms of "classical observable" one $x_{cl}$.

%Note,  in Ref. \cite{bely91-254-153}  the explicit  integration procedure was not presented.  But  the  final EP of order parameter  (Eq.(17))  corresponds to   the factor $(\xi - 3)$, which was used in all  calculations fulfilled therein. But according to  Eq.\Ref{deltaPL}  the opposite  signs of $I_1$ and $I_2$  are impossible.

%Actually, to get the  correct results in Ref. \cite{bely91-254-153}, we have to replace the parameter $(\xi - 3)$ by $(\xi + 1) $ in all the %expressions.
 In particular, the relation between $ x$  and $x_{cl}$ looks as follows
 %(compare with  Eq.(13) in Ref. \cite{bely91-254-153}):
%
\be \label{xclas} x = x_{cl} + \frac{g^2}{4 \pi^2} B_1 (\frac{x_{cl}}{2}) ( \xi + 1)  .\ee
 Within Nielsen's identity approach, this formula  corresponds to  the choice in eq.\Ref{character}  $x' = x_{cl}$  and $\zeta =  - 1$.  Along this orbit the EP  is gauge-fixing independent and  expressed in terms of  $\left< L \right>$. In such a way these two methods are related.

 Inserting eq.\Ref{xclas} in eq.\Ref{EP} and expanding $B_4(\frac{x}{2})$ in powers of $g^2$,  we obtain $ W_L(x_{cl})= W^{(1)}_L(x_{cl})+ W^{(2)}_L(x_{cl}),$ where the first term is obtained from $W^{(1)}(x)$   by means of  substitution $ x \to x_{cl}$ and the second is
\be \label{EPL} \beta^4  W^{(2)}_L(x_{cl}) =  \frac{g^2}{2} \bigl [  B_2^2(\frac{x_{cl}}{2})+ 2 B_2(0) B_2(\frac{x_{cl}}{2}) +
 \frac{8}{3} B_3(\frac{x_{cl}}{2}) B_1(\frac{x_{cl}}{2})\bigr].\ee
In the $W_L(x_{cl})$ the  $\xi$-dependent terms are mutually cancelled, as it should be and demonstrate gauge-fixing independence.

 We also  note that the final  expression for $W_L(x_{cl})$    can be obtained from $W(x)$  eq.\Ref{EP} formally (omitting described consequent  steps)  by means of the next substitutions: $x \to x_{cl}$ and $\xi \to \zeta = - 1$.
 As a result, according eq.\Ref{W0} we  get for the minimum values
\bea \label{Wf} \beta^4  W_L(x_{cl})|_{min}&=&  \beta^4 W_L (0)
-\frac{1}{48 \pi^2} g^4, \nn \\
x_{cl}|_{min}&=&  \frac{g^2}{2 \pi^2} .\eea
Thus, the EP  $W_L(x_{cl})$ has a nonzero minimum position  and does not depend on  $\xi$. The condensation happens at  the two-loop  level. The minimum  value of PL (corresponding to the physical  states)  equals to: $\left< L \right> = \cos( \frac{ g^2}{4 \pi}).$   In contrast,  in \cite{bely91-254-153} the  value  $\left< L \right>  = \pm 1$ was obtained.

The expression $-  W_L(x_{cl})|_{min} = p$ eq.\Ref{Wf}  describes  thermodynamical pressure in the plasma. The first  term is $  \beta^4 W_L (0) = - 0.657974 + \frac{g^2}{24}$.  The function $ W_L(x_{cl})$ can be used for calculating Debye's mass of neutral gluons defined as
\be \label{Dm} m^2_D = \frac{d^2 W_L(x_{cl}) }{d A_0^2}|_{A_0 = 0}, \ee
remind, $x_{cl} = \frac{g A_0^{cl}}{\pi T}.$ We get
\be \label{Dmgl} m^2_D = \frac{2}{3} g^2 T^2 + \frac{5}{4} \frac{g^4}{\pi^2}  T^2. \ee
Here, first term is well known one-loop contribution and the second one is two-loop correction.

To complete we note that the $A_0$ condensation is derived within the  correlation of the one- and two-loop  effective potentials. Whereas asymptotic freedom at high temperature is realized due to the  correlation of the tree-level  and one-loop contributions to the EP.  Formally (as it is often doing in the literature), the latter can be summarized  by the replacement of coupling constant  $g^2 \to \bar{g}^2 \sim \frac{g^2 }{\log(T/T_0)}$, $T_0$ is a reference temperature.  In both cases, the ratio of the relevant terms  is $\sim \frac{g^2}{4 \pi^2}$.  Hence  at high temperature  we can substitute $g^2 \to \bar{g}^2$ in  above formulas, in particular, in eq. \Ref{Wf}.

Thus, the value of the order parameter PL in the minimum of the EP is
\be \label{PLg} \left< L \right> = \cos( \frac{ \bar{g}^2}{4 \pi}).\ee
It gives a possibility for detecting  the deconfinement phase transition and its type. Accounting for the explicit expression for the one-loop effective coupling $\alpha_s = \frac{ \bar{g}^2}{4 \pi}$ in the $SU(2)$ case
\be \label{alpha} \bar{\alpha}_s = \frac{\alpha_s}{1 + \frac{11}{3 \pi} \alpha_s \log(T/T_0)}\ee
we see that the PL is continuously decreasing with  temperature lowering and becomes zero at $ \frac{ \bar{g}^2}{4 \pi} = \frac{\pi}{2}$. This signals  confinement. If we set $T_d = T_0$ the value of the ratio $W^{(2)}/W^{(1)} $ is $\sim 1/2$, that is in the range of applicability of perturbation theory. The phase transition is second order, as it is well known for SU(2) gauge group. We note once again that due to the smallness of the expansion parameter $\frac{g^2}{8 \pi}$ our perturbation  EP of order parameter is suitable instrument for investigating the confinement-deconfinement  phase transition.

An important observation, as we have seen, in order to obtain a $\xi$ independent EP expressed in terms of PL it is sufficient to set $\xi = - 1$ in  expressions of interest. This systematically  will be used in what follows.
\section{\label{T4}The  magnetic field at high temperature}

Let us consider the case of chromomagnetic field at high temperature \cite{skal22-30(1)-3}  and use  eq.(59) of \cite{bord22-82-390}.
 The EP reads
\eq{3.6}{W^{SU(2)}_{gl}&=
	 \frac{b^2}{2 g^2}-\frac{\pi ^2 T^4}{15}
	 -\frac{a_1 b^{3/2} T}{2 \pi }
	 +\frac{11 b^2 \log (4 \pi  T/\mu)}{24 \pi ^2}
	 +g^2\left( \frac{T^4}{24}-\frac{{a_2} \sqrt{b} T^3}{12 \pi }
	 +\frac{{a_2}^2 b T^2}{32 \pi ^2}
	 \right).
}
The first term is energy of classical field. The   terms proportional to $T^4$ constitute the gluon black body radiation. The contribution from the second loop is in the parenthesis. It has ~ $T^3$ behaviour. The numbers $a_1 = 0.828, a_2 = 1.856$ are calculated in \cite{bord22-82-390}, eq.(22). Note that the one-loop part is $\xi-$indepedent and we set $\xi = - 1$ in the two-loop part in order to get gauge invariant EP.

In one-loop order, the energy eq.\Ref{3.6} has a non-trivial minimum resulting from the term proportional to $b^{3/2}T$.   The condensate and the minimum EP  are
\eq{3.7}{ b_{min}^{one} &= \frac{9 {a_1}^2 \alpha_s^2 T^2}{16 \pi ^2},
	&W^{SU(2), \,one}_{min}&=-\frac{\pi ^2 T^4}{15}	-\frac{27 a_1^4  \alpha_s^3 T^4}{512 \pi ^4},
}
where $\alpha_s = g^2/(1 + \frac{11}{12} \frac{g^2}{\pi^2} \log (4 \pi  T/\mu))$ is running coupling constant, $\mu $ is a normalization point for temperature.
The first term of the energy is the gluon black  body radiation. In this approximation, the condensate is  always  present, and the energy in the minimum is always negative. That means the spontaneous vacuum magnetization and color  SU(2) symmetry breaking. Here also an imaginary term presents, but we consider   the real part. The standard way to remove the imaginary term of one-loop effective potential is adding the daisy diagram contributions (see \cite{skal00-576-430} for details). From eq.\Ref{3.7} we see that the presence of $\alpha_s $ weakens the field strength at high temperature.

Now we turn to the two loop case. We consider the high temperature limit and take into consideration the $\sim T^3$ term in eq.\Ref{3.6}. Denoting here $b^{1/2} = x$, we obtain the third-order polynomial equation for determining the  condensate value:
\eq{3.8}{ x^3 - \frac{3}{4 \pi} a_1 T \alpha_s x^2 - \frac{g^2}{2 4 \pi} a_2 T^3 \alpha_s = 0. }
The real root of it can be found using formulas from the  standard handbook \cite{abra64}, Chapter 3.8. The result is
\eq{3.9}{ x_0 = b^{1/2}_{min} = \frac{1}{4}\frac{(2 a_2 \alpha_s)^{1/3}}{\pi^{1/3}} T + \frac{1}{4 \pi} a_1 \alpha_s T .}
If we compare this  with eq.\Ref{3.7}, we find that the second term is three times less than the one in eq.\Ref{3.7}. The most interesting is the change of the temperature dependence coming from $ \alpha_s^{1/3}$. Hence, the first term is dominant for this case.   For the field strength we get in this limit
\eq{3.10}{ b_{min} = \frac{1}{16}\frac{(2 a_2 \alpha_s)^{2/3}}{\pi^{2/3}} T^2.}
Note also, the value $a_2$ is larger than $a_1$. As a result, the role  of the second  loop is important.  Formula  eq.\Ref{3.9} was  derived first in the literature
 in  \cite{skal22-30(1)-3}.

%\begin{figure}[h]%211105_SU(2) a=0.nb
%	\includegraphics[width=0.5\textwidth]{211105_SU2_1.pdf}
%	\includegraphics[width=0.5\textwidth]{211105_SU2_2.pdf}
%	\caption{The  the condensate field and the vacuum energy   in one-loop order   for $SU(2)$ as function of the coupling $g$ for $T=10$ (left panel) and as function of the temperature for $g=1$ (right panel).
%	}
%x	\label{fig:4}\end{figure}
%
\section{\label{T4a}The minimum of the effective potential for pure $A_0$ }
In this section, we remind the known results for the case of a pure $A_0$-background discussed in details in sect. 4.  For $b=0$, the general  effective action eq.\Ref{W2} with  eq.\Ref{1.1} is expressed in terms of Bernoulli's polynomials.  We restrict ourselves to the main topological sector and there to $0\le a\le 1/2$. Here, the EP has a minimum at $a= a_{min}$ (see also eq.(6) in \cite{skal92-7-2895}) and takes in this minimum the value $W_{|{a=a_{min}}}=W_{min}$ with
\eq{4.1}{(g A_0)_{min} &= \frac{3-\xi}{16\pi}g^2 T,
	& W_{min} &= -\frac{\pi^2T^4}{15}
	-\frac{(3-\xi)^2T^4}{192\pi^2}g^4.
}
As mentioned in \cite{skal21-18-738}, \cite{bord21-81-998}, eq.\Ref{4.1} coincides with the gauge-invariant result for $\xi=-1$, what we assume in the following.

Let us compare eq.\Ref{4.1} with the minimal effective potential eq.\Ref{3.7} in the pure magnetic case. We see   in the latter case, the extra temperature dependent factor $(1 + \frac{11}{12} \frac{g^2}{\pi^2} [\log (4 \pi  T)/\mu)^{-1}$ is present and decreases the value of the magnetic condensate at high temperature. For the two loop result eq.\Ref{3.10} the strength of the  field  is larger. But again at sufficiently high temperature the $ \alpha_s^{1/3}$ factor makes the value of $b_{min}(T)$ smaller compared to the value of $(g A_0)_{min}$ eq. \Ref{4.1}.  As a result, since both condensates have negative energies they should  be generated. This decreases the total   free energy of the system. The same takes place for SU(3) gluodynamics and full QCD.
%
%\begin{figure}[h]%211105a_SU(2) a=0.nb
%	\includegraphics[width=0.5\textwidth]{211113_3_1.pdf}
%	\includegraphics[width=0.5\textwidth]{211113_3_2.pdf}
%	\caption{The effective potentials \Ref{3.8} and \Ref{4.1} as functions of the re-scaled variable $a=4 a_{min}s$ and $b=4b_{min}s$  for $T=10$ %(left panel). The  minima of the effective potential at $a=0$, \Ref{3.8}, and at $b=0$. \Ref{4.1},  as function of the coupling $g$ for $T=10$   %(right panel).
%	}
%\label{fig:7}\end{figure}
%

\section{Induced Color charge}
%
%In this section, we calculate the induced color charge generated by the tadpole diagram of Fig. 1. In charged basis, we have two components of %the induced charge for the shifts $A_0^3$ and $A_0^8$. But accounting  for the result \cite{skal21-cond-156} $A_0^8 = 0$ ,  we have to calculate %the contribution for the case $(A_0)_\mu^a=A_0\delta_{\mu4}\delta^{a3}$. The explicit form in the Euclid space-time   is $Q_4^3 Q^3_{ind} $, and %we have
In this section, for QCD, we calculate the induced color charge generated by the tadpole diagram of Fig. 1. In charged basis, we have two components of the induced charge for the shifts $A_0^3$ and $A_0^8$. But accounting  for the result \cite{skal21-cond-156} $A_0^8 = 0$ ,  we have to calculate the contribution for the case $(A_0)_\mu^a=A_0\delta_{\mu4}\delta^{a3}$. The explicit form in the Euclidean  space-time   is $Q_4^3 Q^3_{ind} $, and we have

\begin{equation}
\label{Q3_init}
    Q_{ind}^3=  \frac{g}{\beta}\sum_{p_4}\int\frac{d^3p}{(2\pi)^3}Tr\left[\frac{\lambda^3}{2}\gamma_4\frac{\hat p_\sigma\gamma_\sigma+m}{\hat p^2 +m^2 }\right],
\end{equation}

\noindent where $\hat p=(p_4=p_4\pm A_0,\mathbf{p}),\,p_4=2\pi T(l+1/2),\,l=0,\pm1,\dots$. The trace is calculated over either space-time or color variables, $\lambda^3$ is Gell-Mann matrix.  Here also  we noted as $ A_0 $ the value $ A_0 = \frac{g A_0}{2}$. In what follows we use the Matsubara imaginary time formalism at finite temperature.

Calculating the  traces over the space and the internal indexes we get,

\begin{equation}
\label{DWI_Q3}
    Q_{ind}^3 = \frac{4 g}{\beta} \int\frac{d^3p}{(2\pi)^3}\sum_{p_4}\frac{(p_4+A_0)}{(p_4+A_0)^2 + \ep_p^2},
\end{equation}
where $\ep_p^2 = \vec{p}^2 + m^2$. In case of nonzero field,
$\ep_p^2 = {p_3^2 + m^2 + (2n + 1) gH - g H \sigma} $.  Here we denoted as $g H$ any kind of magnetic field $gH^3, gH^8, eH$ or even some combinations of chromomagnetic fields generated in the plasma at high temperature (see for details \cite{skal18-15-6}).  We have two coupling constants and therefore returned to the H-notations. It is important also that  all the magnetic fields are oriented in one direction in coordinate space. In this case the EP of the fields has minimal energy and so such type ones are generated spontaneously.

%To calculate the temperature sum we use the following representation ($\beta = 1/T$ is inverse temperature)
%
%\begin{equation}
%    Q_{ind}^3= 4 g \int\frac{d^3 p}{(2\pi)^3}\frac{\beta}{\pi} \oint_{C}\tan\frac{\beta\omega}{2}\frac{(\omega+A_0)}{(\omega+A_0)^2+\ep^2_p}d\omega.
%\end{equation}
%
To calculate the temperature sum we use the following representation ($\beta = 1/T$ is inverse temperature)
\begin{equation}
    Q_{ind}^3= 4 g \int\frac{d^3 p}{(2\pi)^3}\frac{\beta}{\pi} \oint_{C}\tan\frac{\beta\omega}{2}\frac{(\omega+A_0)}{(\omega+A_0)^2+\ep^2_p}d\omega.
\end{equation}
The contour $C$ encloses clockwise the poles of tangent in $\omega$-plane.
This is in the case of zero field.
If the field is nonzero, we have to replace $\frac{d^3 p}{(2 \pi)^3} -> \frac{d p_3}{2 \pi}  \frac{gH }{(2 \pi)^2}$ in correspondence to the particle spectrum.  The integrand function has two complex  poles of  first order in the $\omega$-plane. We deform and move the contour to infinity and calculate  the residues of the integrand  to find the charge value.

The result, after transformation into spherical coordinates and angular  integration, is

\begin{equation}
\label{Q1}
    Q_{ind}^3= \frac{g \sin{(A_0 \beta)}}{\pi^2} \int_0^{\infty} p^2 d p \frac{1}{\cos\beta A_0 +\cosh ( \beta \ep_p ) }.
\end{equation}

In what follows, we calculate the integral in the high-temperature limit $T\to\infty$. In this case we use

%\begin{equation}
%    \cep=\sqrt{\mathbf{p}^2+m^2}\approx|\mathbf{p}| + \frac{1}{2}\frac{ m^2}{|\mathbf{p}|}.
%\end{equation}

\begin{equation}
  \ep_p  =\sqrt{\mathbf{p}^2+m^2}\approx|\mathbf{p}| + \frac{1}{2}\frac{ m^2}{|\mathbf{p}|}
\end{equation}
because large values of momentum give dominant contribution.

 After integration over momentum we obtain  at zero field \cite{skal21-cond-156}
\begin{equation}\label{Q3_fin}
    Q_{ind}^3= g A_0^3 \bigl[ \frac{T^2}{3} - \frac{m^2}{2 \pi^2}\bigr].
\end{equation}
As we see,  the first term  depends on temperature as $\sim T^2$. The second one depends on mass, only.  At high-temperature,  the first term is dominant and   plasma acquires the spontaneous induced charge in the case $m=0$, also.

Now, we turn to nonzero $H$.  Using the low Landau level approximation, $\sigma = + 1, n = 0$, we get after integration over $p_3$ momentum
\begin{equation} \label{QHT} Q_{ind}^3(H, T) = g \frac{g H}{2 \pi^3} \frac{\sin (A_0^3 \beta)}{\beta } ( 1 + 7 \beta^2 m^2 Zeta^{'} (- 2) ).
\end{equation}
Note that numerically $Zeta^{'} (- 2) = - 0.03044485$.
Thus, one of the consequences of the $A_0$ condensate presence is the $Z(3)$ symmetry and the $C$-parity violation, which leads to the induction of color charge in the plasma.

%\section{Discussion}
%
Let us compare the values of induced color charges given by formulas eq.(\ref{Q3_fin}) and eq.(\ref{QHT}). The first leading in temperature terms are of interest, now. Both expressions have the factor $g A_0^3$ but a different temperature dependence. In former case, the factor $\sim T^2$ stands and determines high temperature behavior. In latter case, it is determined by the temperature dependence of the magnetic field. This behavior has been investigated in sect. 5  in two-loop approximation eq.\Ref{3.10}.
%\begin{equation}
%\label{bmin} g H^3 (T)  = \frac{1}{16}\frac{(2 a_2 \alpha_s)^{2/3}}{\pi^{2/3}} T^2.
%\end{equation}
%
%Here, $a_2 = 1.856$ is number, $\alpha_s = g^2/(1 + \frac{11}{12} \frac{g^2}{\pi^2} \log (  T/\mu))$ is running coupling constant, $\mu $ is a %normalization point for temperature.  So, at reference temperature $\alpha_s = g^2$.
  Because of the factor $\alpha_s = g^2/(1 + \frac{11}{12} \frac{g^2}{\pi^2} \log (  T/\mu))$, the value of the field strength is always smaller compared to  $T^2$. As a result, the induced color charge eq.\Ref{QHT} in the magnetic field is also smaller compered to eq.\Ref{Q3_fin}.

On the other hand, during carried out calculations we have taken the field strength as a given number which is arbitrary. So that it can be the field produced  by some external current. In this case the induced color charge will be completely determined by external field. Such a situation is expected and discussed for heavy ion collision experiments.

As we noted in Introduction, the factor $g H$ stands for different  magnetic fields - usual magnetic field $eH$, color magnetic fields $gH^3, gH^8$ or even some combination of  them. For instance, in \cite{skal18-15-6} it was  shown that in QGP at the LHC energies the
combinations of fields $H^1_f = q_f H + g (\frac {H^3}{2} + \frac {H^8}{2\sqrt{3}}), H^2_f = q_f H + g (- \frac {H^3}{2} + \frac {H^8}{2\sqrt{3}}), H^3_f = q_f H  - g  \frac {H^8}{2\sqrt{3}}$, have to be spontaneously produced,  where $q_f $ is electric charge of quark species. The strengths  of the fields at various temperatures  have been estimated. In particular, it was shown that the strength of color fields is two order stronger    compared to the usual magnetic one.  The spectra of all charged particles become discrete that influences and specifies   the manifestations of QGP. It is also important   that (as we noted above) all the generated fields  are collinearly directed  in space.
% This  is because  for such  orientation  the free energy is lower.

It is also interesting  that the field presence  decreases  the phase transition temperature. This also has been obtained in analytic  \cite{skal18-15-6} and  lattice computations \cite{demc08-41-165051}, \cite{cea99-60-094506}, \cite{cea05-0508-079}. As a general conclusion of above consideration, the most important consequence of the induced charge $Q^3_{ind.}$ is the generation of classical static corol potential $q^3_{stat.}$ that opens a possibility for new scattering processes in the QGP. The magnetic fields modify them in an essential way. So, numerous manifestations of the plasma creation could be observed in  experiments. For example, these influence the number of direct photons radiated from QGP, modify scattering of photons on it, etc.  These phenomena will be discussed elsewhere.
\section{Effective $\gamma \gamma G $  vertexes in $QGP$}
Other interesting objects which have to be generated in the $QGP$
with $A_0$ condensate are the effective three-line vertexes
$\gamma \gamma G^3, \gamma \gamma G^8. $ They  should exist
because of Furry's theorem violation and relate color  and white
states.   These   vertexes, in particular,  have to result in
observable processes of new type - inelastic scattering of
photons, splitting (dissociation / conversion) of gluon
$\bar{\phi}^3$, $\bar{\phi}^8$ classical potentials in two photons. As we have shown above, $Q^8_{ind}$ is not generated in the considered approximation. But it is not excluded in higher-loop orders. So we have to take in mind the second type  vertex also.
%\begin{figure}[h!]
%\centering
%\includegraphics[scale=0.5]{loop-1ext}
%\hspace{40pt}
%\includegraphics[scale=0.5]{loop-3ext}
%\hspace{40pt}
%\includegraphics[scale=0.5]{loop-pol-op}
%\end{figure}

In this  section, we calculate the vertex $\gamma \gamma
G^3$ depicted in Fig. 2  and investigate some related processes in the plasma.
%These can be  signals of the creation of  QGP.
%\begin{figure}[h!]
%\centering
%\includegraphics{loop-gm-gm-gluon}
%\end{figure}
%
%Let us introduce the notations used in what follows.
The   vertex $\Gamma_{\mu \la}^\nu$  consists of two such type diagrams. The
second one is obtained by changing the direction of the quark
line. We use the notations: all the momenta are ingoing, first
photon $\gamma_1 (k^1_\mu)$, second photon $\gamma_2 (k^3_\la)$,
color a=3 gluon $Q^3 (k^2_\nu) $, and $k^1 + k^2 + k^3 = 0$.
$k^{1,2,3}$ are momenta of external fields.

We consider the contributions coming from the traces of four
$\gamma$-matrixes, which are   proportional to the quark mass and
dominant for small photon momenta $k^1, k^3 << m$. The analytic
expression reads
\be \label{gamma}  \Gamma_{\mu \la}^\nu (k^1, k^3) =  \Gamma_{\mu
\la}^{\nu, (1)}(k^1, k^3) +   \Gamma_{\mu \la}^{\nu, (2)}(k^1,
k^3), \ee
where
\be \label{gamma1}  \Gamma_{\mu \la}^{\nu, (1)}(k^1, k^3) =
\frac{1}{\beta} \sum_{p_4} \int \frac{d^3 p}{(2 \pi)^3}
\frac{N_{1}}{D(\tilde{P}) D(\tilde{P}- k^1) D(\tilde{P} +
k^3)}.\ee
Here $\beta = T^{-1}$, summation is over
$p_4 = 2\pi T ( l + 1/2), l = 0, \pm 1, \pm 2, ...$,  integration
is over three dimensional momentum space $p$, $N_1$ denotes the
numerator coming from   the first diagram, $\tilde{P} =
(\tilde{P}_4 = p_4 - A_0 , \vec{p})$, $D(\tilde{P}) = ( p_4 -
A_0)^2 + \vec{p}^2 + m^2  = \tilde{P}_4 ^2 +  \epsilon_p^2 $ and $
\epsilon_p^2 = \vec{p}^2 + m^2$ is energy of free quark squared. In case of nonzero field, $\epsilon_p^2 = {p}_3^2 + m^2 + (2 n + 1) g H - g H \sigma$ as in previous section.  We also have to replace $\frac{d^3 p}{(2 \pi)^3} -> \frac{d p_3}{2 \pi}  \frac{gH }{(2 \pi)^2}$.

First we consider the zero field case.
The functions $ D(\tilde{P}- k^1),  D(\tilde{P} + k^3) $ assume a
corresponding shift in momentum. The numerator $N_1$ is
\be \label{N1}( N_1 )_{\mu\nu\la} = \delta_{\mu\nu} (\tilde{P} -
k^2)_\la + \delta_{\la\nu} (\tilde{P} - k^2)_\mu + \delta_{\mu\la}
(\tilde{P} - q)_\nu, \ee
where $ q = k^3 - k^1$ is photon momentum   transferred.

The expression for the second term in eq.\Ref{gamma} is coming from
the second diagram and obtained from  eqs.\Ref{gamma1}, \Ref{N1} by
substitutions $k^1 \to - k^1, k^2 \to - k^2, q \to - q$. We denote
the second numerator as $N_2$. In what follows we carry out actual
calculations for the first term in eq.\Ref{gamma} and adduce the
results for the second one.

Now, we  take into consideration the fact that in the high
temperature limit the large  values of the integration  momentum $p$
give  leading contributions. Therefore we can present the
functions $D(\tilde{P}),  D(\tilde{P}- k^1),  D(\tilde{P} + k^3)$
in the form:
\bea \label{Di}  D(\tilde{P})  & =& \tilde{P}_4 ^2 +  \epsilon_p^2
=  \tilde{P} ^2, \\ \nn
 D(\tilde{P}- k^1) &=&  \tilde{P} ^2 \Bigl(1-  \frac{2  \tilde{P}\cdot k^1 - k_1^2}{\tilde{P} ^2} \Bigr), \\ \nn
   D(\tilde{P}+ k^3) &=&  \tilde{P} ^2 \Bigl( 1+  \frac{2  \tilde{P}\cdot k^3 + k_3^2}{\tilde{P} ^2} \Bigr). \eea
%$
Here, $k_1^2 = (k^1_4)^2 + \vec{k}_1^2,   k_3^2 = (k^3_4)^2 +
\vec{k}_3^2$. At high  temperature and $ \tilde{P} ^2  \to \infty$
the k-dependent terms are  small.   So, we can expand in these
parameters and obtain for the integrand    in eq.\Ref{gamma1}
%
%\bea \label{Intgr1} Integr_1  &&= \frac{N_1}{(\tilde{P} ^2)^3} \Bigl[ 1 - 2 \frac{2 (\tilde{P}\cdot q)}{\tilde{P}^2} - \frac{k^2_3 - k_1^2}{\tilde{P}^2}
%& &`- 4 \frac{(\tilde{P}\cdot k^1) (\tilde{P}\cdot k^3)}{\tilde{P} ^2} + 4  \frac{(\tilde{P}\cdot k^1)^2 + (\tilde{P}\cdot k^3)^2}{\tilde{P} ^2} \Birl]. \eea
%
\be \label{Intd} Intd.  = \frac{N_1}{(\tilde{P} ^2)^3} \Bigl[ 1 +
\sum_{i=1}^4 A_i \Bigr], \ee
where \be \label{Ai} A_1 = - 2 \frac{ (\tilde{P}\cdot
q)}{\tilde{P}^2}, ~~A_2 =  - \frac{k^2_3 - k_1^2}{\tilde{P}^2} \ee

\be \label{Aj}  A_3 = - 4 \frac{(\tilde{P}\cdot k^1)
(\tilde{P}\cdot k^3)}{\tilde{P} ^2}, ~~ A_4 =  4
\frac{(\tilde{P}\cdot k^1)^2 + (\tilde{P}\cdot k^3)^2}{\tilde{P}
^2} \ee
and vector $q_\mu = (q_4, \vec{q})$.

For the  second diagram we have to substitute $q \to - q$,  other
terms are even and do not change.

Further we concentrate on the scattering of photons on the
potential $Q_4^3$ in the  medium rest frame and set the thermostat
velocity $u_\nu = (1, \vec{0}), \nu = 4$. The corresponding terms
in the numerators are
\be \label{N0}  N_1 -> \delta_{\mu\la} (\tilde{P} + q )_4, ~~  N_2
-> \delta_{\mu\la} (\tilde{P} - q )_4, \ee
remind that $\tilde{P}_4 = p_4 - A_0$  and $\tilde{P}^2 = ( p_4 -
A_0)^2 + \epsilon_p^2$. In this case the numerators do not depend on space momentum and therefore also the magnetic field presence. So that we proceed further  with the zero field case and take the field into consideration when it will be necessary.

We have to calculate in general the series of two types
corresponding to these numerators:
\be \label{Ns1} S_1^{(n)} =\frac{1}{\beta} \sum_{p_4} \frac{p_4 -
A_0}{(\tilde{P}^2 )^n},  ~~ S_2^{(n)} =\frac{1}{\beta} \sum_{p_4}
\frac{q_4}{(\tilde{P}^2 )^n}, n = 3, 4, 5. \ee
These functions can be calculated from the
$ S_1^{(1)} $
 and
 $ S_2^{(1)}$
  by computing  a number of derivatives with respect to
$\epsilon_p^2$. The latter series result in simple expressions.
First is the one calculated already  for the tadpole diagram
eq.\Ref{Q1}. But now we have to change the sing $A_0 \to - A_0$.
\be \label{S11} S_1^{(1)} = \frac{1}{\beta} \sum_{p_4} \frac{p_4 -
A_0}{\tilde{P}^2 } = - \frac{1}{2} \frac{\sin(A_0 \beta)}{\cos
(A_0 \beta) + \cosh(\epsilon_p \beta)}. \ee
The function $  S_2^{(1)}$ is
\be \label{21} S_2^{(1)} = \frac{1}{\beta} \sum_{p_4}
\frac{q_4}{\tilde{P}^2 } = - \frac{q_4}{2 \epsilon_p}
\frac{\sinh(\epsilon_p \beta)}{\cos (A_0 \beta) + \cosh(\epsilon_p
\beta)}. \ee
Hence, the explicit analytic expressions can be obtained for the
integrand and the integration  over $d^3 p$ is carried out in terms of
known  functions and their derivatives.

%Before doing that, we present the functions \Ref{Ai}, \Ref{Aj} in
%the form convenient for integration. In particular, the scattering
%angle between $\vec{k}_3 $ and $\vec{k}_1 $  naturally appears  in
%$A_3$.

%To present the kind of algebraic transformations we consider the
%numerator of  $A_3$ term in Eq.\Ref{Aj}
%
%\bea \label{transA3}  (\tilde{P}\cdot k^3)}{\tilde{P} ^2& =&
%(\tilde{P}_4 k^1_4 + \vec{p}\cdot \vec{k}_1 )  (\tilde{P}_4 k^3_4
%+ \vec{p}\cdot \vec{k}_3 )\\ \nn & =&  \tilde{P} ^2  k^1_4 k^3_4 -
%\epsilon^2_p   k^1_4 k^3_4 + \tilde{P}_4 k^3_4 ( \vec{p}\cdot
%\vec{k}_1) + \tilde{P}_4 k^1_4 (\vec{p}\cdot \vec{k}_3) +
%(\vec{p}\cdot \vec{k}_1)  ( \vec{p}\cdot \vec{k}_3). \eea
%
%The $\tilde{P} ^2$  in the first term is canceled with denominator
%and  the linear in $\vec{p}$ terms  can be omitted because the
%corresponding integral over $d^3 p$ is zero. In such a way other
%$A_i$ terms can be transformed.

Let us adduce the  expressions for $A_i$ obtained after some
simplifying algebraic transformations:
 \be \label{A1} A_1 = - 2
\frac{(p_4 - A_0) q_4}{\tilde{P} ^2},\ee
\be \label{A2} A_3 = -\frac{ 4}{\tilde{P} ^2} \Bigl[ (1 -
\frac{\epsilon^2_p}{  \tilde{P} ^2}) k^1_4 k^3_4 +
\frac{(\vec{p}\cdot \vec{k}_1)  ( \vec{p}\cdot \vec{k}_3)}{
\tilde{P} ^2} \Bigr],\ee
\be \label{A4} A_4 =\frac{ 4}{\tilde{P} ^2} \Bigl[ (1 -
\frac{\epsilon^2_p}{  \tilde{P} ^2})(( k^1_4)^2+ ( k^3_4)^2) +
\frac{(\vec{p}\cdot \vec{k}_1)^2 +  ( \vec{p}\cdot \vec{k}_3)^2 }{
\tilde{P} ^2} \Bigr],\ee
Accounting for the structure of the numerators in eq.\Ref{N0}, we see
that  the terms without $\tilde{P}_4$ are canceled in the sum of
two diagrams and the resulting amplitude consists of the expressions
\be \label{M1} M_1 = 2 \delta_{\mu\la} \frac{p_4 - A_0}{(\tilde{P}
^2)^3} ( 1 + A_1 + A_3 + A_4  ) \ee
and \be \label{M2} M_2 = - 4 \delta_{\mu\la} \frac{(p_4 - A_0)
q^2_4}{(\tilde{P} ^2)^4}. \ee
Thus, all the contributions of the $S_2^{(n)}$ series are canceled
in the total.

Now we tern to $d^3 p$ integration.
%
%\section{ Integration in leading order}
%
The expressions in eqs.\Ref{M1}, \Ref{M2} contain different powers
of $\tilde{P} ^2$, and hence different powers of $\beta$ appear
even in the leading $p \to \infty$ approximation, which
corresponds to the first term in the expansion $\epsilon_p = p +
\frac{1}{2} \frac{m^2}{p} + O(p^{-3})$. Below, we carry out
integration in this leading in $T \to \infty$ approximation.
%The next- to-
%leading terms will be considered latter.

We present our procedure considering the first term in eq.\Ref{M1}
which is calculated as the  second derivative of $S_1^{(1)}$ over
$\epsilon_p^2$ and  equaled  to
\be \label{S3} S_3 = - A_0 \beta ~\frac{(Sech(\beta \epsilon_p
/2))^4}{64 p^3} ( - 2 \beta \epsilon_p + \beta \epsilon_p
Cosh(\beta \epsilon_p ) +  Sinh(\beta \epsilon_p ) ). \ee
Then in the spherical coordinates we calculate the integral
\be \label{I3} I_3 = \int\limits_{-\infty}^{\infty} d^3 p~ S_3  =
4 \pi   \int\limits_{0}^{\infty  }  p^2 d p  ~S_3(p). \ee
 In leading order $\epsilon_p \beta = p \beta$.   Making
the change of variables $p \beta = y$ we obtain for eq.\Ref{I3},
\be \label{I3a} I_3 = - \frac{A_0 \pi \beta}{16}
\int\limits_{0}^{\infty  } \frac{d y}{y} (Sech(y /2))^4 ( - 2 y + y
Cosh(y ) +  Sinh(y ) ). \ee
Note that this integral is convergent.
% The $S_3$  small $y$
%expansion is
%
%\be \label{S30} S_3 (y \to 0) = - A_0 \beta^4 ( \frac{1}{96}  +
%\frac{17 y^2}{3840}). \ee
%
Numeric integration in eq.\Ref{I3a} gives
\be \label{I3num} I_3 = - A_0 \pi \beta ~(0.3348). \ee
In such a way all the other integrations in eqs.\Ref{M1}, \Ref{M2}
can be carried out.
% We see that each the integral carries own power of
%$\beta$. Hence actual temperature dependence of amplitude is
%presented.

Performing analogous calculations for other terms we obtain the
  expression for scattering amplitude in high
temperature approximation. A not complicated problem is to find
next-to-leading corrections having the order $(m \beta)^l, l = 1,
2, ...$ As a result, the explicit high temperature limits  for the
scattering amplitude can be calculated in terms of elementary
functions.

Now, we consider the case of non zero magnetic fields. As mentioned before, the expression eq.\Ref{S3} independs of the magnetic field presence and does not change. In the lower level approximation $n = 0, \sigma = 1$ we have $\epsilon_p^2=p_3^2 + m^2$. So, for large momenta $\epsilon_p = p_3 (1 + \frac{1}{2}\frac{p_3}{m})$. The expression eq.\Ref{I3} now has the form
\be \label{I3h} I_3 = \int\limits_{-\infty}^{\infty} \frac{d^3 p}{(2 \pi)^3} ~ S_3  =
\int\limits_{-\infty}^{\infty} \frac{d p_3}{2 \pi}   \frac{gH}{(2\pi)^2}  ~S_3(p_3). \ee
This integrand function is even with respect to the change $p_3 \to - p_3$, so we can calculate two integrals in the limits $(0 , \infty)$.
Performing numeric integrations we obtain,
\be \label{I3Hnum} I_3 (H)  = gH  A_0  \beta^3 ~(0.0283). \ee
Here, for magnetic field $gH$ different variants can be substituted, as it is noted in previous sections. The most important conclusion is that in magnetized QGP   various processes generated by these effective three-linear photon-photon-gluon vertexes have to happen. They should serve as the signals of the deconfinement phase transition. Since magnetic fields depend on temperature in the described above way, corresponding  numeric estimates (and numbers) could be obtained for investigated processes.

Among interesting processes is the conversion   of classical static  gluon fields $ \bar{\phi}^3(k),  \bar{\phi}^8(k)$   (generated in the plasma due to the  color charges $Q^3_{ind.}, Q^8_{ind.})$, in photons. This conversion formally looks as a super radiance in condense matter physics (because of static initial states).
Due to   the effective
vertex $ \Gamma^\nu_{\mu\la}(k^1, k^3)$, in the rest frame of the
plasma two photons  moving in opposite directions and having
specific energies, which  correspond to quantum or classical states,  have to be radiated. In fact, one has to detect a classical photon flow corresponding to the decaying classical state.

 One of the consequences of this effect has to be an increase   of known  in the literature direct infrared  photons radiated from the QGP in  heavy ion collisions. That acts to remove  the deficit of direct low frequency photons existing in the known calculations compared to the experimental data. Independently of the process resulting in direct photons, there exists a cut in photon frequency which can be radiated from the plasma. The $A_0$ condensate lowers this frequency and in this way increases in theory the yield of direct low frequency photons. More details see, for instance, in \cite{sinykov} and references therein. The magnetic fields modify the vertex $ \Gamma^\nu_{\mu\la}(k^1, k^3)$ that also influences the output of direct photons as well.

These processes  could serve as the signals of the QGP creation.

\end{document}